\documentclass[reprint,aps,pra,amsmath,amssymb,showpacs,nofootinbib]{revtex4-1}
\usepackage{xcolor}
\usepackage{hyperref}
\hypersetup{
	breaklinks=true,
	colorlinks=true,
	linkcolor=blue,
	citecolor=purple,
	pdfstartview={FitH},
	pdfview={FitH 0},%pdfstartview=FitH,pdfview=FitH,%pdfstartview={XYZ null null 1},
	pdfauthor={Jeffrey B. Parker},
 }

\usepackage{bm}
\usepackage{eucal}		% better fonts for \mathcal
\usepackage{mycommands}   % load after hyperref

\providecommand{\eqnref}{}
\renewcommand{\eqnref}[1]{Eq.~\eqref{#1}}

\usepackage{braket}

\begin{document}

\title{Quantum phase estimation for a class of generalized eigenvalue problems}

\author{Jeffrey B.~Parker}
\email{jeff.parker@wisc.edu}
\affiliation{Department of Physics, University of Wisconsin-Madison, Madison, WI 53706, USA}
\affiliation{Lawrence Livermore National Laboratory, Livermore, CA 94550, USA}

\author{Ilon Joseph}
\affiliation{Lawrence Livermore National Laboratory, Livermore, CA 94550, USA}

\begin{abstract}
Quantum phase estimation provides a path to quantum computation of solutions to Hermitian eigenvalue problems $Hv = \l v$, such as those occurring in quantum chemistry.  It is natural to ask whether the same technique can be applied to generalized eigenvalue problems $Av = \l B v$, which arise in many areas of science and engineering.  We answer this question affirmatively.  A restricted class of generalized eigenvalue problems could be solved as efficiently as standard eigenvalue problems.  A paradigmatic example is provided by Sturm--Liouville problems.  Another example comes from linear ideal magnetohydrodynamics, where phase estimation could be used to determine the stability of magnetically confined plasmas in fusion reactors. 
\end{abstract}

\maketitle

\section{Introduction}
Quantum computing has been recognized for its potential to solve currently intractable physics problems, such as quantum many-body problems \cite{lloyd:1996}.  The ability to process information in a fundamentally quantum way is also being explored for physics problems more broadly.  For example, a quantum algorithm for the wave equation has been recently proposed \cite{costa:2019}.  Montanaro and Pallister proposed a quantum method of solving boundary-value problems via finite-element discretization \cite{montanaro:2016}.  A quantum algorithm for linear initial-value differential equations has been proposed by Berry \emph{et al.}~\cite{berry:2014,berry:2017}.  Engel \emph{et al.}~proposed a quantum algorithm for the Vlasov equation \cite{engel:2019}. Quantum linear system solvers are invoked in many of these algorithms \cite{harrow:2009,childs:2017}.

Another problem that arises frequently in physics and engineering is computation of spectra.  It is often the case that a few eigenvalues are of primary interest, such as the calculation of the ground state and first few excited states.  A related example is the determination of whether a system is linearly stable, and if unstable, what the growth rate is.

Quantum phase estimation (QPE) is a core quantum algorithm that offers quantum speedups for this problem.  Fundamentally, QPE provides a technique for finding eigenvalues of unitary matrices.    Moreover, QPE forms an important quantum subroutine in other algorithms, such as Shor's algorithm \cite{shor:1994} and the Harrow--Hassidim--Lloyd algorithm \cite{harrow:2009}.  Abrams and Lloyd \cite{abrams:1999} realized that QPE could be used to calculate eigenvalues and eigenstates of Hamiltonians, in effect providing a quantum algorithm to solve the eigenvalue problem $H v = \l v$.  QPE could be used to find, for example, the electronic ground state in atomic physics or quantum chemistry problems \cite{mcardle:2018arxiv}.  In such problems that stem from quantum physics, the number of degrees in the Schr\"{o}dinger equation increases exponentially with the number of particles, and thus a numerical solution is intractable via classical computation for many particles.  As a result, QPE may become a key algorithm for solving ground states in a future error corrected, digital quantum computer.

In this paper, we show that QPE can also be usefully applied for calculating eigenvalues and eigenvectors of a certain class of \emph{generalized} eigenvalue problems that arise naturally in physical applications.  Generalized eigenvalue problems take the form $Av  = \l Bv$ where $A$ and $B$ are matrices.  We assume that $A$ is Hermitian and $B$ is positive definite.  Examples within this restricted class include regular Sturm--Liouville problems.  Another important example is linear ideal magnetohydrodynamics (MHD), which is an important model for determining stability of magnetically confined plasmas such as those used in fusion reactors.  We demonstrate that generalized eigenvalue problems within the restricted class may be reduced to a standard eigenvalue problem in such a way that the effective Hamiltonian is still sparse, and thus potentially efficiently implementable.  Then one can apply the standard procedure of Abrams and Lloyd \cite{abrams:1999}.

Our motivating problems of interest arise from partial differential equations in classical physics, such as continuum dynamics, rather than quantum problems.  These problems typically involve polynomial growth with respect to a parameter such as the number of grid points, rather than exponential growth.  For instance, if some physical quantity is represented on a $d$-dimensional grid with $N$ points per dimension, the total number of grid points required is $N^d$.  Yet these problems can pose great challenges in three dimensions in complex geometry when extremely high resolution is required.  Computing eigenvalues classically requires $O[\text{poly}(N^d)]$ arithmetic operations.  For grid-based discretizations of partial differential equations, it has not been established that an exponential speedup is possible; estimates have given a possible polynomial speedup \cite{szkopek:2005}.

An outline for the rest of the paper is as follows.  Section~\ref{sec:phaseestimation} provides a brief overview of quantum phase estimation.  In Sec.~\ref{sec:generalized_evp}, we describe how QPE can be applied to a restricted class of generalized eigenvalue problems.  In Sec.~\ref{sec:sturmliouville}, we apply this idea to the Sturm--Liouville problem and discuss different discretizations.  In Sec.~\ref{sec:idealmhd}, we describe a problem of some practical importance to which this technique could be applied, the determination of stability of magnetized plasmas.  Finally, we give our concluding remarks in Sec.~\ref{sec:discussion}.

\section{Quantum Phase Estimation}
\label{sec:phaseestimation}
We recall the main facets of quantum phase estimation.  Phase estimation is a technique that, given an eigenvector $v$ of a unitary matrix $U$, can provide the eigenvalue.  Since the eigenvalues of a unitary matrix have magnitude 1, the eigenvalue is often written as $e^{2\pi i \p}$ with $0 \le \p < 1$.  

QPE can be applied to solving for eigenvalues of a Hermitian eigenvalue problem:
	\begin{equation}
		H v = \l v.
	\end{equation}
For the moment, we assume the unitary gate $U_H = e^{-iHt}$ can be efficiently applied.  For example, assume $0 \le \l < 1$ and let $t = 2\pi$.  If the desired eigenvalue $\l$ is not between 0 and 1, $H$ can be shifted by a multiple of identity and scaled by modifying $t$ to bring the eigenvalue into this range.  If $v$ is an eigenvector of $H$ with eigenvalue $\l$, then $v$ is also an eigenvector of $U_H$, with $U_H v = e^{-2\pi i \l} v$.  

The critical components of the algorithm are state preparation, Hamiltonian simulation, and measurement of the phase.  We briefly discuss each of these aspects.

State preparation is an important part of phase estimation.  The desired eigenfunction is typically not known in advance, and the input state must be sufficiently close to an eigenvector.  If the normalized eigenvectors of $H$ are $v_j$, and the input state is $v$, the probability of determining the eigenvalue of state $j$ is the overlap $P_j = | \langle v |v_j \rangle |^2$, and thus to measure a specific eigenvalue $\l_j$ requires on average $1/P_j$ trials.  We will not delve into state preparation in any further depth here.  We merely remark that the best way to approximate the desired eigenvector is problem dependent.  Knowledge of the specific physics of the problem can often be used to vastly narrow the space of interest.  For instance, the WKB method can provide first approximations to solutions of partial differential equations. Classically computed approximations on a coarser grid may be an alternative approach \cite{jaksch:2003}.  Adiabatic evolution is another method by which to prepare states \cite{aspuruguzik:2005}.

The textbook approach to phase estimation uses the inverse quantum Fourier transform, where each bit of precision requires an additional ancilla qubit \cite{nielsen:book}.   Because quantum computers in the near term will have constrained numbers of qubits, there is active research into variants of phase estimation to improve the practical performance.  The Kitaev method, for instance, requires only one additional qubit \cite{kitaev:book}.  Other QPE variations offer reductions in the required circuit depth, coherence time, and number of samples \cite{svore:2014, obrien:2019}.  Additionally, algorithms resilient to error from experimental imperfections have been explored \cite{wiebe:2016, dobsicek:2007}.

Error estimates for phase estimation have been discussed elsewhere \cite{abrams:1999,szkopek:2005}.  One aspect that merits further comment is whether exponential speedup is attainable for the partial differential equations that serve as our motivation.  So far, it has not been established that any discretization based on a real-space uniform grid can lead to exponential speedup \cite{szkopek:2005}.  At most polynomial speedup has been shown in a fixed number of dimensions.  The problem is not unique to phase estimation but arises with any Hamiltonian simulation \cite{kivlichan:2017}.  The standard error estimate leads to the spectral norm problem, which is that the error associated with digital quantum simulation of a Hamiltonian $H$ scales with the norm $|H|$.

To briefly show how this error estimate arises, for simplicity consider simulation based on first-order Trotterization of $H = H_1 + H_2$.  Simulation for a time $t$ in $m$ Trotter steps incurs an error $\ve$, which can be estimated using the Baker--Campbell--Hausdorff formula to be \cite{childs:thesis}
	\begin{equation}
		\ve = O\left( \frac{|[H_1, H_2]| t^2}{m} \right).
		\label{errortotal:standardestimate}
	\end{equation}

For concreteness, now consider the one-dimensional Schr\"{o}dinger equation where $H_1 = -d^2/dx^2$ and $H_2 = V(x)$, discretized on a spatial grid with $N$ grid points and fixed domain size.  As one refines the grid by increasing $N$, the norm of the discrete differential operator grows as the inverse of the grid spacing.  For a second derivative in $H_1$, one has $|H_1| \sim N^2$, and one can estimate
	\begin{equation}
		|[H_1, H_2]| \sim N^2.
		\label{H1H2norm_generic}
	\end{equation}
To simulate for a given time $t$ with constant error, it appears that $m$ must grow as $N^2$, which is exponential in the number of qubits $n = \log N$.  Thus, Hamiltonian simulation may require so many Trotter steps to ensure accuracy that there is no exponential speedup \cite{szkopek:2005,somma:2016}.  Other algorithms for Hamiltonian simulation besides Trotterization produce similar conclusions; Ref.~\cite{berry:2015} provides an algorithm with complexity scaling in the worst case as $|H|_{\text{max}}$, the largest entry of $H$ in absolute value.

\section{Generalized Eigenvalue Problem}
\label{sec:generalized_evp}
Consider the generalized eigenvalue problem
	\begin{equation}
		A v = \l B v.
		\label{GEP}
	\end{equation}
Although $\l$ and $v$ are sometimes referred to as generalized eigenvalues and generalized eigenvectors, we follow a typical physics convention of simply calling them eigenvalues and eigenvectors.  We assume $A$ is Hermitian, $A_{ij} = A_{ji}^*$.  We also restrict to considering cases where $B$ is positive definite.  In general, $A$ and $B$ need not commute.  Under these conditions, just like Hermitian eigenvalue problems, the eigenvalues are real, and eigenvectors with distinct eigenvalues are orthogonal when weighted by $B$.  That is, if $v_1$ and $v_2$ are eigenvectors with different eigenvalues, then $v_1^\dagger B v_2 = 0$.

%\subsection{Quantum Algorithm}
Since quantum phase estimation can compute eigenvalues for the standard eigenvalue problem, it is natural to ask whether the same technique can be applied to the generalized eigenvalue problem.  We answer this question affirmatively: for at least some cases, generalized eigenvalue problems can be solved as efficiently as standard eigenvalue problems.  Since our primary motivation comes from discretization of differential equations, we think of $A$ as arising from a differential operator, and therefore we assume $A$ is local and sparse.  Typically $A$ will have a banded structure.  In many problems of interest, $B$ will also be sparse.  

In considering \eqnref{GEP}, even if we assume we can apply $U_A=e^{-iAt}$ and $U_B = e^{-iBt}$ efficiently, it does not appear the techniques of QPE can straightforwardly provide a solution to yield the eigenvalues.  That is because QPE relies on the fact that $H$ or $U_H$ applied to an eigenvector $v$ produces a multiple of $v$.  In the generalized eigenvalue problem, it is not true that applying $U_A$ to $v$ produces a multiple of $v$, as can be seen from \eqnref{GEP}.

One could rewrite \eqnref{GEP} as $B^{-1} A v = \l v$, however one runs into the immediate problem that the matrix $B^{-1} H$ in general will not be Hermitian.  Instead we use the fact that a positive-definite matrix $B$ has a unique positive-definite square root $B^{1/2}$.  We introduce the change of variables
	\begin{equation}
		u \defineas B^{1/2} v,
	\end{equation}
which transforms \eref{GEP} into
	\begin{equation}
		\widetilde{H} u \defineas B^{-1/2} A B^{-1/2} u = \l u,
		\label{GEP_transformed}
	\end{equation}
and $\widetilde{H}$ is Hermitian.  The problem has thus been reduced to standard Hermitian eigenvalue form to which QPE could in principle be applied.

A potential issue with \eqnref{GEP_transformed} is that in general, even if $B$ is sparse, $B^{-1/2}$ is not sparse and $\widetilde{H}$ will be full.  For example, the inverse of a tridiagonal matrix is dense.  When $\widetilde{H}$ is full and in the absence of some other special structure, it is not likely that the unitary gate $U_{\widetilde{H}}$ required for QPE can be implemented efficiently.

%  In the context of generalized eigenvalue problems and applying QPE, this transformation has been noted before in work on dimensionality reduction.  

For a restricted class of matrices $B$, however, $B^{-1/2}$ will be sparse.  In particular, if in addition to being positive definite $B$ is also diagonal or block diagonal, its inverse retains the same sparseness pattern, and therefore $\widetilde{H}$ will also be sparse.  In the case where $B$ is diagonal, $B_{ij} = b_i \de_{ij}$, then $\widetilde{H}_{ij} = b_i^{-1/2} A_{ij} b_j^{-1/2}$.  Therefore, $U_{\widetilde{H}} = e^{-i \widetilde{H} t}$ could be efficiently implementable as a unitary gate, enabling phase estimation as described in Sec.~\ref{sec:phaseestimation} for this class of generalized eigenvalue problems.

An alternative route to reduce the generalized eigenvalue problem into a standard one uses the Cholesky decomposition $B = L L^\dagger$ where $L$ is lower triangular.  The new change of variables
	\begin{equation}
		w \defineas L^\dagger v
	\end{equation}
transforms \eref{GEP} into
	\begin{equation}
		\overline{H} w \defineas L^{-1} A (L^\dagger)^{-1} w = \l w,
	\end{equation}
where $\overline{H}$ is Hermitian.  Again, when $B$ is block diagonal, $L^{-1}$ is sparse.

It is interesting to compare the sparseness of $\widetilde{H}$, resulting from the $B^{1/2}$ decomposition, and that of $\overline{H}$, resulting from the Cholesky decomposition.  To be specific, we adopt the following parameters for this comparison: let $A$ be banded with $2k + 1$ nonzero bands and let $B$ be block diagonal with blocks of size $m$.  The matrices $A$ and $B$ have size $N \times N$.  For large $N$, the number of nonzero elements of $\overline{H}$ is approximately $(2k + m)N$.  Similarly, the number of nonzero elements of $\widetilde{H}$ in the case $m \ge k$ is approximately $3mN$.  Thus, we see that $\overline{H}$ is more sparse than $\widetilde{H}$.  For example, suppose $A$ is tridiagonal with $k=1$ and that the blocks of $B$ are of size $m=4$.  The number of nonzero entries of $\overline{H}$ is half that of $\widetilde{H}$.  This greater sparseness using the Cholesky decomposition may enable more efficient solutions on a quantum computer.

%Classical algorithms will all require a number of operations the scale polynomially in $N^d$. 
%To compare with the classical algorithm,... to find all eigenvalues: $QZ$ decomposition (refs... Golub and Van Loan).  To find one eigenvalue for sparse matrices: Arnoldi iteration (refs).  Time scaling.  (Also do not need to know the eigenvector for this algorithm to work)

\section{Sturm--Liouville Problem}
\label{sec:sturmliouville}
\subsection{Finite-difference discretization}
The Sturm--Liouville problem presents a paradigmatic example of a generalized eigenvalue problem.  It takes the form
	\begin{equation}
		-\d{}{x} \left( p(x) \d{y}{x} \right) + q(x) y(x) = \l r(x) y(x),
		\label{SLproblem}
	\end{equation}
where $\l$ is the eigenvalue, and for boundary conditions we assume $y(0) = y(1) = 0$.  We assume a regular Sturm--Liouville problem, for which $p>0$ and $r>0$ on $[0, 1]$.

Quantum phase estimation for a restricted form of the Sturm--Liouville equation has been discussed in Ref.~\cite{papageorgiou:2005}.  That work, however, set $r(x)=1$, and therefore was already in standard eigenvalue form rather than a generalized eigenvalue problem.  A nonconstant weight function $r(x)$ leads to a generalized eigenvalue problem.

To solve this equation numerically, some kind of discretization is needed.  We could discretize first to obtain a generalized eigenvalue matrix equation and then apply the transformation in Eq.~\eref{GEP_transformed}.  However, it is instructive to apply the analogous transform to the continuous equation first and then discretize.  In Eq.~\eref{SLproblem}, we set $u = r^{1/2} y$, obtaining the equivalent problem 
	\begin{equation}
	\frac{1}{r^{1/2}} \left( -\d{}{x} p \d{}{x} + q \right) \frac{1}{r^{1/2}} u = \l u,		
		%-\frac{1}{r^{1/2}} \d{}{x} \left[ p \d{}{x} \left( \frac{u}{r^{1/2}} \right) \right] + \frac{q}{r} u = \l u.
		\label{SLproblem_transformed}
	\end{equation}

Suppose we use a finite-difference discretization.  Consider a grid with equally spaced points where $x_j = j \D x$ and $\D x = 1/(n+1)$.  The unknown function $u$ is replaced with a vector of $u$ evaluated at grid points $u_j = u(x_j)$ at $j=1, \ldots, n$ (boundary conditions mean the $u_0$ and $u_{n+1}$ are known).  A finite-difference discretization of the left-hand side of \eqnref{SLproblem_transformed} is
		%-\frac{1}{r^{1/2}} \d{}{x} \left( p(x) \d{}{x} \frac{u}{r^{1/2}} \right) \to  \widetilde{H} \begin{bmatrix} u_1 \\ u_2 \\ u_3 \\ \vdots \end{bmatrix},
	\begin{equation}
		\frac{1}{r^{1/2}} \left( -\d{}{x}  p \d{}{x} + q \right) \frac{1}{r^{1/2}} u  \to  \widetilde{H} \begin{bmatrix} u_1 \\ u_2 \\ u_3 \\ \vdots \end{bmatrix},
	\end{equation}
where $\widetilde{H}$ is the tridiagonal matrix
	\begin{widetext}
	\begin{equation}
		\widetilde{H} = \frac{1}{\D x^2} \begin{bmatrix}
		r_1^{-1} \bigl[-(p_{1/2} + p_{3/2}) + q_1 \D x^2\bigr]		&		r_1^{-1/2} p_{3/2} r_2^{-1/2} & & & \\
		r_1^{-1/2} p_{3/2} r_2^{-1/2}	&		r_2^{-1} \bigl[-(p_{3/2} + p_{5/2}) + q_2 \D x^2 \bigr]		&	r_2^{-1/2} p_{5/2} r_3^{-1/2} & & \\
															&		r_2^{-1/2} p_{5/2} r_3^{-1/2}	&	r_3^{-1} \bigl[-(p_{5/2} + p_{7/2}) + q_3 \D x^2 \bigr]	  & r_3^{-1/2} p_{7/2} r_4^{-1/2} & \\
															&															&		&		\ddots & 
		\end{bmatrix}.
		\label{sturmliouville:Htilde}
	\end{equation}
	\end{widetext}
Here, $p_j = p(x_j)$, $r_j = r(x_j)$, and unwritten entries in the matrix are zero.  QPE can now be applied directly.

A natural higher-dimensional generalization of the Sturm--Liouville problem involves the Laplace--Beltrami operator.  We consider a generalized eigenvalue problem
	\begin{equation}
		\left[\pd{}{x^i} J(x) g^{ij}(x) \pd{}{x^j} + J(x)V(x) \right] y = \l J r(x) y,
	\end{equation}
where now $x=\bigl(x^1, \ldots, x^m \bigr)$ is an $m$-dimensional coordinate, $J(x)$ is the Jacobian, and $g$ is the metric.  If $r$ is positive definite, one can apply the same transformation $u = (Jr)^{1/2} y$ and obtain a Hermitian eigenvalue problem.

\subsection{Effect of finite-element basis}
The Sturm--Liouville example employed a finite-difference discretization with a position basis, in which case the weight function $r(x)$ turned into a diagonal mass matrix.  Let us consider other discretizations, such as the finite element method.  In a Galerkin procedure using a set of basis functions $\p_i$, the matrix $B$ arises as
	\begin{equation}
		B_{ij} = \int dx\, r(x) \p_i^*(x) \p_j(x)
		\label{finiteelement_Bmatrix}
	\end{equation}
Some classes of finite-element methods, including discontinuous Galerkin, can turn the weight function into a block-diagonal mass matrix.  This can occur when a discretization uses disjoint cells, with each cell spanned by multiple basis functions, which may be orthogonal or nonorthogonal.  If the basis functions within a cell are nonorthogonal, the matrix $B$ is block diagonal rather than diagonal.  In fact, even if $r(x)=1$ and there is no weight function, this choice of basis functions will lead to block-diagonal positive-definite $B$.  When $B$ is block diagonal, so is $B^{-1}$, and hence $B^{-1}$ is sparse.  Then one can find $B^{-1/2}$ which is also block diagonal.  Therefore, $B^{-1/2} A B^{-1/2}$ will be sparse as well, allowing its time evolution operator to be implemented efficiently.

In contrast, other common finite-element discretizations may not lead to sparse $B^{-1/2}$.  For example, consider the linear tent functions $\p_j$ in one dimension, where each basis function has finite support and overlaps with each of its two neighbors.  The matrix $B$ is then tridiagonal.  For this discretization, $B^{-1/2}$ and $\widetilde{H}$ will not be sparse.

\section{Linear Ideal MHD}
\label{sec:idealmhd}
Another important example of a nontrivial generalized eigenvalue problem to which phase estimation might be fruitfully applied arises from plasma physics.  MHD describes the self-consistent motion of a conducting fluid with a magnetic field.  MHD can describe the macroscopic behavior of a plasma, at a fluid level that coarse grains over the motion of individual particles, and it has found applications in stellar physics, astrophysical disks, and planetary magnetism.  When describing magnetically confined plasmas, such as those used in tokamak fusion reactors, ideal MHD has proven essential for calculations of macroscopic plasma stability \cite{troyon:1984,connor:1998,ryutov:2011}.

In linear ideal MHD, the task of solving normal-mode stability of a static plasma equilibrium takes the form of a generalized eigenvalue problem.  The normal-mode equation is
	\begin{equation}
		\v{F}\bigl(\bm{\xi}(\v{x}) \bigr) = -\w^2 \r(\v{x}) \bm{\xi}(\v{x}),
		\label{ideal_mhd_eqn}
	\end{equation}
where $\bm{\xi}$ is the Lagrangian fluid displacement, $\w^2$ is the eigenvalue, $\r(\v{x})$ is the equilibrium mass density of the MHD fluid, and $\v{F}$ is the force operator, given by \cite{bernstein:1958}
	\begin{align}
		\v{F}(\bm{\xi}) &= -\nabla p_1  -  \v{B}_0 \times (\nabla \times \v{B}_1)  \notag \\
		& \quad + (\nabla \times \v{B}_0) \times \v{B}_1  + (\nabla \Phi) \nabla \cdot (\r \bm{\xi}). 
	\end{align}
Here, $\Phi$ is an external potential, and $p_1$ and $\v{B}_1$ are the perturbed pressure and magnetic field and are linear in $\bm{\xi}$.  Here, $\v{F}$ is a differential operator, and therefore local.

It is a fundamental result of MHD theory that $\v{F}$ is a self-adjoint operator \cite{bernstein:1958}.  That is, given any two allowed displacements $\bm{\xi}$ and $\bm{\eta}$,
	\begin{equation}
		\int d\v{x}\, \bm{\eta}^* \cdot \v{F}( \bm{\xi} )  =  \int d\v{x}\, \bm{\xi}^* \cdot \v{F}( \bm{\eta}).
		\label{idealmhd_selfadjointness}
	\end{equation}
From the self-adjointness property, it follows that $\w^2$ is real.  One has normal-mode instability if there are any modes for which $\w^2 < 0$, which corresponds to an imaginary frequency and an exponentially growing solution.  

Discretization of \eqnref{ideal_mhd_eqn} in the position basis, e.g., through finite-difference discretization, 
yields a sparse Hermitian matrix on the left-hand side and a diagonal, positive matrix on the right-hand side.  Thus, the method described in Sec.~\ref{sec:generalized_evp} is applicable.  There are subtleties associated with singularities of the MHD spectrum, which are beyond the scope of this paper, although methods have been developed to deal with them in classical computation \cite{cheng:1987}.

\section{Conclusion}
\label{sec:discussion}
We have shown that quantum phase estimation can be applied to certain generalized eigenvalue problems of the form $Av = \l B v$, where $A$ is a Hermitian matrix and $B$ is a positive-definite matrix.  This is accomplished by reducing this problem to a standard eigenvalue problem by defining $\widetilde{H} = B^{-1/2} A B^{-1/2}$.  For this reduction to lead to a sparse Hamiltonian whose time evolution can be implemented efficiently, $B^{-1/2}$ must be sparse.  For example, $B$ can be diagonal or block diagonal.  An alternative reduction based on the Cholesky decomposition instead of the square root was also discussed.  Many classes of physics problems arise where the fundamental continuous equations, when discretized, produce a diagonal or block-diagonal matrix $B$ that is positive definite.  Some examples are Sturm--Liouville eigenvalue problems and the normal-mode formulation of the spectral equation in ideal MHD.

Even for systems whose continuous formulation is in standard Hermitian eigenvalue form, such as the time-independent Schr\"{o}dinger equation, some choices of nonorthogonal basis functions lead to a generalized eigenvalue problem where $B$ is block diagonal.  The techniques described here can again be used to reduce the problem to standard form on which phase estimation can be applied.

\vspace{4cm}
\begin{acknowledgments}
We gratefully acknowledge discussions with J. DuBois, V. Geyko, F. Graziani, and Y. Shi.  This work was performed under the auspices of the U.S.\ Department of Energy (DOE) by Lawrence Livermore National Laboratory under Contract No.\ DE-AC52-07NA27344 and was supported by the U.S. DOE Office of Fusion Energy Sciences, ``Quantum Leap for Fusion Energy Sciences'' (FWP Grant No.\ SCW1680).
\end{acknowledgments}
 
%\bibliographystyle{apsrev4-2}
%\bibliography{gen_eig_refs}	

\begin{thebibliography}{30}%
\makeatletter
\providecommand \@ifxundefined [1]{%
 \@ifx{#1\undefined}
}%
\providecommand \@ifnum [1]{%
 \ifnum #1\expandafter \@firstoftwo
 \else \expandafter \@secondoftwo
 \fi
}%
\providecommand \@ifx [1]{%
 \ifx #1\expandafter \@firstoftwo
 \else \expandafter \@secondoftwo
 \fi
}%
\providecommand \natexlab [1]{#1}%
\providecommand \enquote  [1]{``#1''}%
\providecommand \bibnamefont  [1]{#1}%
\providecommand \bibfnamefont [1]{#1}%
\providecommand \citenamefont [1]{#1}%
\providecommand \href@noop [0]{\@secondoftwo}%
\providecommand \href [0]{\begingroup \@sanitize@url \@href}%
\providecommand \@href[1]{\@@startlink{#1}\@@href}%
\providecommand \@@href[1]{\endgroup#1\@@endlink}%
\providecommand \@sanitize@url [0]{\catcode `\\12\catcode `\$12\catcode
  `\&12\catcode `\#12\catcode `\^12\catcode `\_12\catcode `\%12\relax}%
\providecommand \@@startlink[1]{}%
\providecommand \@@endlink[0]{}%
\providecommand \url  [0]{\begingroup\@sanitize@url \@url }%
\providecommand \@url [1]{\endgroup\@href {#1}{\urlprefix }}%
\providecommand \urlprefix  [0]{URL }%
\providecommand \Eprint [0]{\href }%
\providecommand \doibase [0]{https://doi.org/}%
\providecommand \selectlanguage [0]{\@gobble}%
\providecommand \bibinfo  [0]{\@secondoftwo}%
\providecommand \bibfield  [0]{\@secondoftwo}%
\providecommand \translation [1]{[#1]}%
\providecommand \BibitemOpen [0]{}%
\providecommand \bibitemStop [0]{}%
\providecommand \bibitemNoStop [0]{.\EOS\space}%
\providecommand \EOS [0]{\spacefactor3000\relax}%
\providecommand \BibitemShut  [1]{\csname bibitem#1\endcsname}%
\let\auto@bib@innerbib\@empty
%</preamble>
\bibitem [{\citenamefont {Lloyd}(1996)}]{lloyd:1996}%
  \BibitemOpen
  \bibfield  {author} {\bibinfo {author} {\bibfnamefont {S.}~\bibnamefont
  {Lloyd}},\ }\bibfield  {title} {\bibinfo {title} {Universal quantum
  simulators},\ }\href {https://doi.org/10.1126/science.273.5278.1073}
  {\bibfield  {journal} {\bibinfo  {journal} {Science}\ }\textbf {\bibinfo
  {volume} {273}},\ \bibinfo {pages} {1073} (\bibinfo {year}
  {1996})}\BibitemShut {NoStop}%
\bibitem [{\citenamefont {Costa}\ \emph {et~al.}(2019)\citenamefont {Costa},
  \citenamefont {Jordan},\ and\ \citenamefont {Ostrander}}]{costa:2019}%
  \BibitemOpen
  \bibfield  {author} {\bibinfo {author} {\bibfnamefont {P.~C.~S.}\
  \bibnamefont {Costa}}, \bibinfo {author} {\bibfnamefont {S.}~\bibnamefont
  {Jordan}},\ and\ \bibinfo {author} {\bibfnamefont {A.}~\bibnamefont
  {Ostrander}},\ }\bibfield  {title} {\bibinfo {title} {Quantum algorithm for
  simulating the wave equation},\ }\href
  {https://doi.org/10.1103/PhysRevA.99.012323} {\bibfield  {journal} {\bibinfo
  {journal} {Phys. Rev. A}\ }\textbf {\bibinfo {volume} {99}},\ \bibinfo
  {pages} {012323} (\bibinfo {year} {2019})}\BibitemShut {NoStop}%
\bibitem [{\citenamefont {Montanaro}\ and\ \citenamefont
  {Pallister}(2016)}]{montanaro:2016}%
  \BibitemOpen
  \bibfield  {author} {\bibinfo {author} {\bibfnamefont {A.}~\bibnamefont
  {Montanaro}}\ and\ \bibinfo {author} {\bibfnamefont {S.}~\bibnamefont
  {Pallister}},\ }\bibfield  {title} {\bibinfo {title} {Quantum algorithms and
  the finite element method},\ }\href
  {https://doi.org/10.1103/PhysRevA.93.032324} {\bibfield  {journal} {\bibinfo
  {journal} {Phys. Rev. A}\ }\textbf {\bibinfo {volume} {93}},\ \bibinfo
  {pages} {032324} (\bibinfo {year} {2016})}\BibitemShut {NoStop}%
\bibitem [{\citenamefont {Berry}(2014)}]{berry:2014}%
  \BibitemOpen
  \bibfield  {author} {\bibinfo {author} {\bibfnamefont {D.~W.}\ \bibnamefont
  {Berry}},\ }\bibfield  {title} {\bibinfo {title} {High-order quantum
  algorithm for solving linear differential equations},\ }\href
  {https://doi.org/10.1088/1751-8113/47/10/105301} {\bibfield  {journal}
  {\bibinfo  {journal} {J. Phys. A: Math. Theor.}\ }\textbf {\bibinfo {volume}
  {47}},\ \bibinfo {pages} {105301} (\bibinfo {year} {2014})}\BibitemShut
  {NoStop}%
\bibitem [{\citenamefont {Berry}\ \emph {et~al.}(2017)\citenamefont {Berry},
  \citenamefont {Childs}, \citenamefont {Ostrander},\ and\ \citenamefont
  {Wang}}]{berry:2017}%
  \BibitemOpen
  \bibfield  {author} {\bibinfo {author} {\bibfnamefont {D.~W.}\ \bibnamefont
  {Berry}}, \bibinfo {author} {\bibfnamefont {A.~M.}\ \bibnamefont {Childs}},
  \bibinfo {author} {\bibfnamefont {A.}~\bibnamefont {Ostrander}},\ and\
  \bibinfo {author} {\bibfnamefont {G.}~\bibnamefont {Wang}},\ }\bibfield
  {title} {\bibinfo {title} {Quantum algorithm for linear differential
  equations with exponentially improved dependence on precision},\ }\href
  {https://doi.org/10.1007/s00220-017-3002-y} {\bibfield  {journal} {\bibinfo
  {journal} {Commun. Math. Phys.}\ }\textbf {\bibinfo {volume} {356}},\
  \bibinfo {pages} {1057} (\bibinfo {year} {2017})}\BibitemShut {NoStop}%
\bibitem [{\citenamefont {Engel}\ \emph {et~al.}(2019)\citenamefont {Engel},
  \citenamefont {Smith},\ and\ \citenamefont {Parker}}]{engel:2019}%
  \BibitemOpen
  \bibfield  {author} {\bibinfo {author} {\bibfnamefont {A.}~\bibnamefont
  {Engel}}, \bibinfo {author} {\bibfnamefont {G.}~\bibnamefont {Smith}},\ and\
  \bibinfo {author} {\bibfnamefont {S.~E.}\ \bibnamefont {Parker}},\ }\bibfield
   {title} {\bibinfo {title} {Quantum algorithm for the vlasov equation},\
  }\href {https://doi.org/10.1103/PhysRevA.100.062315} {\bibfield  {journal}
  {\bibinfo  {journal} {Phys. Rev. A}\ }\textbf {\bibinfo {volume} {100}},\
  \bibinfo {pages} {062315} (\bibinfo {year} {2019})}\BibitemShut {NoStop}%
\bibitem [{\citenamefont {Harrow}\ \emph {et~al.}(2009)\citenamefont {Harrow},
  \citenamefont {Hassidim},\ and\ \citenamefont {Lloyd}}]{harrow:2009}%
  \BibitemOpen
  \bibfield  {author} {\bibinfo {author} {\bibfnamefont {A.~W.}\ \bibnamefont
  {Harrow}}, \bibinfo {author} {\bibfnamefont {A.}~\bibnamefont {Hassidim}},\
  and\ \bibinfo {author} {\bibfnamefont {S.}~\bibnamefont {Lloyd}},\ }\bibfield
   {title} {\bibinfo {title} {Quantum algorithm for linear systems of
  equations},\ }\href {https://doi.org/10.1103/PhysRevLett.103.150502}
  {\bibfield  {journal} {\bibinfo  {journal} {Phys. Rev. Lett.}\ }\textbf
  {\bibinfo {volume} {103}},\ \bibinfo {pages} {150502} (\bibinfo {year}
  {2009})}\BibitemShut {NoStop}%
\bibitem [{\citenamefont {Childs}\ \emph {et~al.}(2017)\citenamefont {Childs},
  \citenamefont {Kothari},\ and\ \citenamefont {Somma}}]{childs:2017}%
  \BibitemOpen
  \bibfield  {author} {\bibinfo {author} {\bibfnamefont {A.~M.}\ \bibnamefont
  {Childs}}, \bibinfo {author} {\bibfnamefont {R.}~\bibnamefont {Kothari}},\
  and\ \bibinfo {author} {\bibfnamefont {R.~D.}\ \bibnamefont {Somma}},\
  }\bibfield  {title} {\bibinfo {title} {Quantum algorithm for systems of
  linear equations with exponentially improved dependence on precision},\
  }\href {https://doi.org/10.1137/16M1087072} {\bibfield  {journal} {\bibinfo
  {journal} {SIAM J. Comput.}\ }\textbf {\bibinfo {volume} {46}},\ \bibinfo
  {pages} {1920} (\bibinfo {year} {2017})}\BibitemShut {NoStop}%
\bibitem [{\citenamefont {{Shor}}(1994)}]{shor:1994}%
  \BibitemOpen
  \bibfield  {author} {\bibinfo {author} {\bibfnamefont {P.~W.}\ \bibnamefont
  {{Shor}}},\ }\bibfield  {title} {\bibinfo {title} {Algorithms for quantum
  computation: discrete logarithms and factoring},\ }in\ \href
  {https://doi.org/10.1109/SFCS.1994.365700} {\emph {\bibinfo {booktitle}
  {Proceedings 35th Annual Symposium on Foundations of Computer Science}}}\
  (\bibinfo {year} {1994})\ pp.\ \bibinfo {pages} {124--134}\BibitemShut
  {NoStop}%
\bibitem [{\citenamefont {Abrams}\ and\ \citenamefont
  {Lloyd}(1999)}]{abrams:1999}%
  \BibitemOpen
  \bibfield  {author} {\bibinfo {author} {\bibfnamefont {D.~S.}\ \bibnamefont
  {Abrams}}\ and\ \bibinfo {author} {\bibfnamefont {S.}~\bibnamefont {Lloyd}},\
  }\bibfield  {title} {\bibinfo {title} {Quantum algorithm providing
  exponential speed increase for finding eigenvalues and eigenvectors},\ }\href
  {https://doi.org/10.1103/PhysRevLett.83.5162} {\bibfield  {journal} {\bibinfo
   {journal} {Phys. Rev. Lett.}\ }\textbf {\bibinfo {volume} {83}},\ \bibinfo
  {pages} {5162} (\bibinfo {year} {1999})}\BibitemShut {NoStop}%
\bibitem [{\citenamefont {McArdle}\ \emph {et~al.}(2018)\citenamefont
  {McArdle}, \citenamefont {Endo}, \citenamefont {Aspuru-Guzik}, \citenamefont
  {Benjamin},\ and\ \citenamefont {Yuan}}]{mcardle:2018arxiv}%
  \BibitemOpen
  \bibfield  {author} {\bibinfo {author} {\bibfnamefont {S.}~\bibnamefont
  {McArdle}}, \bibinfo {author} {\bibfnamefont {S.}~\bibnamefont {Endo}},
  \bibinfo {author} {\bibfnamefont {A.}~\bibnamefont {Aspuru-Guzik}}, \bibinfo
  {author} {\bibfnamefont {S.}~\bibnamefont {Benjamin}},\ and\ \bibinfo
  {author} {\bibfnamefont {X.}~\bibnamefont {Yuan}},\ }\href@noop {} {\bibinfo
  {title} {Quantum computational chemistry}} (\bibinfo {year} {2018}),\ \Eprint
  {https://arxiv.org/abs/1808.10402} {arXiv:1808.10402 [quant-ph]} \BibitemShut
  {NoStop}%
\bibitem [{\citenamefont {Szkopek}\ \emph {et~al.}(2005)\citenamefont
  {Szkopek}, \citenamefont {Roychowdhury}, \citenamefont {Yablonovitch},\ and\
  \citenamefont {Abrams}}]{szkopek:2005}%
  \BibitemOpen
  \bibfield  {author} {\bibinfo {author} {\bibfnamefont {T.}~\bibnamefont
  {Szkopek}}, \bibinfo {author} {\bibfnamefont {V.}~\bibnamefont
  {Roychowdhury}}, \bibinfo {author} {\bibfnamefont {E.}~\bibnamefont
  {Yablonovitch}},\ and\ \bibinfo {author} {\bibfnamefont {D.~S.}\ \bibnamefont
  {Abrams}},\ }\bibfield  {title} {\bibinfo {title} {Eigenvalue estimation of
  differential operators with a quantum algorithm},\ }\href
  {https://doi.org/10.1103/PhysRevA.72.062318} {\bibfield  {journal} {\bibinfo
  {journal} {Phys. Rev. A}\ }\textbf {\bibinfo {volume} {72}},\ \bibinfo
  {pages} {062318} (\bibinfo {year} {2005})}\BibitemShut {NoStop}%
\bibitem [{\citenamefont {Jaksch}\ and\ \citenamefont
  {Papageorgiou}(2003)}]{jaksch:2003}%
  \BibitemOpen
  \bibfield  {author} {\bibinfo {author} {\bibfnamefont {P.}~\bibnamefont
  {Jaksch}}\ and\ \bibinfo {author} {\bibfnamefont {A.}~\bibnamefont
  {Papageorgiou}},\ }\bibfield  {title} {\bibinfo {title} {Eigenvector
  approximation leading to exponential speedup of quantum eigenvalue
  calculation},\ }\href {https://doi.org/10.1103/PhysRevLett.91.257902}
  {\bibfield  {journal} {\bibinfo  {journal} {Phys. Rev. Lett.}\ }\textbf
  {\bibinfo {volume} {91}},\ \bibinfo {pages} {257902} (\bibinfo {year}
  {2003})}\BibitemShut {NoStop}%
\bibitem [{\citenamefont {Aspuru-Guzik}\ \emph {et~al.}(2005)\citenamefont
  {Aspuru-Guzik}, \citenamefont {Dutoi}, \citenamefont {Love},\ and\
  \citenamefont {Head-Gordon}}]{aspuruguzik:2005}%
  \BibitemOpen
  \bibfield  {author} {\bibinfo {author} {\bibfnamefont {A.}~\bibnamefont
  {Aspuru-Guzik}}, \bibinfo {author} {\bibfnamefont {A.~D.}\ \bibnamefont
  {Dutoi}}, \bibinfo {author} {\bibfnamefont {P.~J.}\ \bibnamefont {Love}},\
  and\ \bibinfo {author} {\bibfnamefont {M.}~\bibnamefont {Head-Gordon}},\
  }\bibfield  {title} {\bibinfo {title} {Simulated quantum computation of
  molecular energies},\ }\href {https://doi.org/10.1126/science.1113479}
  {\bibfield  {journal} {\bibinfo  {journal} {Science}\ }\textbf {\bibinfo
  {volume} {309}},\ \bibinfo {pages} {1704} (\bibinfo {year}
  {2005})}\BibitemShut {NoStop}%
\bibitem [{\citenamefont {Nielsen}\ and\ \citenamefont
  {Chuang}(2010)}]{nielsen:book}%
  \BibitemOpen
  \bibfield  {author} {\bibinfo {author} {\bibfnamefont {M.~A.}\ \bibnamefont
  {Nielsen}}\ and\ \bibinfo {author} {\bibfnamefont {I.~L.}\ \bibnamefont
  {Chuang}},\ }\href {https://doi.org/10.1017/CBO9780511976667} {\emph
  {\bibinfo {title} {Quantum Computation and Quantum Information: 10th
  Anniversary Edition}}}\ (\bibinfo  {publisher} {Cambridge University Press},\
  \bibinfo {year} {2010})\BibitemShut {NoStop}%
\bibitem [{\citenamefont {Kitaev}\ \emph {et~al.}(2002)\citenamefont {Kitaev},
  \citenamefont {Shen},\ and\ \citenamefont {Vyalyi}}]{kitaev:book}%
  \BibitemOpen
  \bibfield  {author} {\bibinfo {author} {\bibfnamefont {A.~Y.}\ \bibnamefont
  {Kitaev}}, \bibinfo {author} {\bibfnamefont {A.~H.}\ \bibnamefont {Shen}},\
  and\ \bibinfo {author} {\bibfnamefont {M.~N.}\ \bibnamefont {Vyalyi}},\
  }\href@noop {} {\emph {\bibinfo {title} {Classical and Quantum
  Computation}}}\ (\bibinfo  {publisher} {American Mathematical Society},\
  \bibinfo {address} {USA},\ \bibinfo {year} {2002})\BibitemShut {NoStop}%
\bibitem [{\citenamefont {Svore}\ \emph {et~al.}(2014)\citenamefont {Svore},
  \citenamefont {Hastings},\ and\ \citenamefont {Freedman}}]{svore:2014}%
  \BibitemOpen
  \bibfield  {author} {\bibinfo {author} {\bibfnamefont {K.~M.}\ \bibnamefont
  {Svore}}, \bibinfo {author} {\bibfnamefont {M.~B.}\ \bibnamefont
  {Hastings}},\ and\ \bibinfo {author} {\bibfnamefont {M.}~\bibnamefont
  {Freedman}},\ }\bibfield  {title} {\bibinfo {title} {Faster phase
  estimation},\ }\href@noop {} {\bibfield  {journal} {\bibinfo  {journal}
  {Quantum Info. Comput.}\ }\textbf {\bibinfo {volume} {14}},\ \bibinfo {pages}
  {306–328} (\bibinfo {year} {2014})}\BibitemShut {NoStop}%
\bibitem [{\citenamefont {O'Brien}\ \emph {et~al.}(2019)\citenamefont
  {O'Brien}, \citenamefont {Tarasinski},\ and\ \citenamefont
  {Terhal}}]{obrien:2019}%
  \BibitemOpen
  \bibfield  {author} {\bibinfo {author} {\bibfnamefont {T.~E.}\ \bibnamefont
  {O'Brien}}, \bibinfo {author} {\bibfnamefont {B.}~\bibnamefont
  {Tarasinski}},\ and\ \bibinfo {author} {\bibfnamefont {B.~M.}\ \bibnamefont
  {Terhal}},\ }\bibfield  {title} {\bibinfo {title} {Quantum phase estimation
  of multiple eigenvalues for small-scale (noisy) experiments},\ }\href
  {https://doi.org/10.1088/1367-2630/aafb8e} {\bibfield  {journal} {\bibinfo
  {journal} {New J. Phys.}\ }\textbf {\bibinfo {volume} {21}},\ \bibinfo
  {pages} {023022} (\bibinfo {year} {2019})}\BibitemShut {NoStop}%
\bibitem [{\citenamefont {Wiebe}\ and\ \citenamefont
  {Granade}(2016)}]{wiebe:2016}%
  \BibitemOpen
  \bibfield  {author} {\bibinfo {author} {\bibfnamefont {N.}~\bibnamefont
  {Wiebe}}\ and\ \bibinfo {author} {\bibfnamefont {C.}~\bibnamefont
  {Granade}},\ }\bibfield  {title} {\bibinfo {title} {Efficient bayesian phase
  estimation},\ }\href {https://doi.org/10.1103/PhysRevLett.117.010503}
  {\bibfield  {journal} {\bibinfo  {journal} {Phys. Rev. Lett.}\ }\textbf
  {\bibinfo {volume} {117}},\ \bibinfo {pages} {010503} (\bibinfo {year}
  {2016})}\BibitemShut {NoStop}%
\bibitem [{\citenamefont {Dobsicek}\ \emph {et~al.}(2007)\citenamefont
  {Dobsicek}, \citenamefont {Johansson}, \citenamefont {Shumeiko},\ and\
  \citenamefont {Wendin}}]{dobsicek:2007}%
  \BibitemOpen
  \bibfield  {author} {\bibinfo {author} {\bibfnamefont {M.}~\bibnamefont
  {Dobsicek}}, \bibinfo {author} {\bibfnamefont {G.}~\bibnamefont {Johansson}},
  \bibinfo {author} {\bibfnamefont {V.}~\bibnamefont {Shumeiko}},\ and\
  \bibinfo {author} {\bibfnamefont {G.}~\bibnamefont {Wendin}},\ }\bibfield
  {title} {\bibinfo {title} {Arbitrary accuracy iterative quantum phase
  estimation algorithm using a single ancillary qubit: A two-qubit benchmark},\
  }\href {https://doi.org/10.1103/PhysRevA.76.030306} {\bibfield  {journal}
  {\bibinfo  {journal} {Phys. Rev. A}\ }\textbf {\bibinfo {volume} {76}},\
  \bibinfo {pages} {030306} (\bibinfo {year} {2007})}\BibitemShut {NoStop}%
\bibitem [{\citenamefont {Kivlichan}\ \emph {et~al.}(2017)\citenamefont
  {Kivlichan}, \citenamefont {Wiebe}, \citenamefont {Babbush},\ and\
  \citenamefont {Aspuru-Guzik}}]{kivlichan:2017}%
  \BibitemOpen
  \bibfield  {author} {\bibinfo {author} {\bibfnamefont {I.~D.}\ \bibnamefont
  {Kivlichan}}, \bibinfo {author} {\bibfnamefont {N.}~\bibnamefont {Wiebe}},
  \bibinfo {author} {\bibfnamefont {R.}~\bibnamefont {Babbush}},\ and\ \bibinfo
  {author} {\bibfnamefont {A.}~\bibnamefont {Aspuru-Guzik}},\ }\bibfield
  {title} {\bibinfo {title} {Bounding the costs of quantum simulation of
  many-body physics in real space},\ }\href
  {https://doi.org/10.1088/1751-8121/aa77b8} {\bibfield  {journal} {\bibinfo
  {journal} {J. Phys. A: Math. Theor.}\ }\textbf {\bibinfo {volume} {50}},\
  \bibinfo {pages} {305301} (\bibinfo {year} {2017})}\BibitemShut {NoStop}%
\bibitem [{\citenamefont {Childs}(2004)}]{childs:thesis}%
  \BibitemOpen
  \bibfield  {author} {\bibinfo {author} {\bibfnamefont {A.}~\bibnamefont
  {Childs}},\ }\emph {\bibinfo {title} {Quantum information processing in
  continuous time}},\ \href@noop {} {Ph.D. thesis} (\bibinfo {year}
  {2004})\BibitemShut {NoStop}%
\bibitem [{\citenamefont {Somma}(2016)}]{somma:2016}%
  \BibitemOpen
  \bibfield  {author} {\bibinfo {author} {\bibfnamefont {R.~D.}\ \bibnamefont
  {Somma}},\ }\bibfield  {title} {\bibinfo {title} {Quantum simulations of one
  dimensional quantum systems},\ }\href@noop {} {\bibfield  {journal} {\bibinfo
   {journal} {Quantum Info. Comput.}\ }\textbf {\bibinfo {volume} {16}},\
  \bibinfo {pages} {1125–1168} (\bibinfo {year} {2016})}\BibitemShut
  {NoStop}%
\bibitem [{\citenamefont {{Berry}}\ \emph {et~al.}(2015)\citenamefont
  {{Berry}}, \citenamefont {{Childs}},\ and\ \citenamefont
  {{Kothari}}}]{berry:2015}%
  \BibitemOpen
  \bibfield  {author} {\bibinfo {author} {\bibfnamefont {D.~W.}\ \bibnamefont
  {{Berry}}}, \bibinfo {author} {\bibfnamefont {A.~M.}\ \bibnamefont
  {{Childs}}},\ and\ \bibinfo {author} {\bibfnamefont {R.}~\bibnamefont
  {{Kothari}}},\ }\bibfield  {title} {\bibinfo {title} {Hamiltonian simulation
  with nearly optimal dependence on all parameters},\ }in\ \href
  {https://doi.org/10.1109/FOCS.2015.54} {\emph {\bibinfo {booktitle} {2015
  IEEE 56th Annual Symposium on Foundations of Computer Science}}}\ (\bibinfo
  {year} {2015})\ pp.\ \bibinfo {pages} {792--809}\BibitemShut {NoStop}%
\bibitem [{\citenamefont {Papageorgiou}\ and\ \citenamefont
  {Wo{\'{z}}niakowski}(2005)}]{papageorgiou:2005}%
  \BibitemOpen
  \bibfield  {author} {\bibinfo {author} {\bibfnamefont {A.}~\bibnamefont
  {Papageorgiou}}\ and\ \bibinfo {author} {\bibfnamefont {H.}~\bibnamefont
  {Wo{\'{z}}niakowski}},\ }\bibfield  {title} {\bibinfo {title} {Classical and
  quantum complexity of the sturm--liouville eigenvalue problem},\ }\href
  {https://doi.org/10.1007/s11128-005-4481-x} {\bibfield  {journal} {\bibinfo
  {journal} {Quantum Inf. Process.}\ }\textbf {\bibinfo {volume} {4}},\
  \bibinfo {pages} {87} (\bibinfo {year} {2005})}\BibitemShut {NoStop}%
\bibitem [{\citenamefont {Troyon}\ \emph {et~al.}(1984)\citenamefont {Troyon},
  \citenamefont {Gruber}, \citenamefont {Saurenmann}, \citenamefont
  {Semenzato},\ and\ \citenamefont {Succi}}]{troyon:1984}%
  \BibitemOpen
  \bibfield  {author} {\bibinfo {author} {\bibfnamefont {F.}~\bibnamefont
  {Troyon}}, \bibinfo {author} {\bibfnamefont {R.}~\bibnamefont {Gruber}},
  \bibinfo {author} {\bibfnamefont {H.}~\bibnamefont {Saurenmann}}, \bibinfo
  {author} {\bibfnamefont {S.}~\bibnamefont {Semenzato}},\ and\ \bibinfo
  {author} {\bibfnamefont {S.}~\bibnamefont {Succi}},\ }\bibfield  {title}
  {\bibinfo {title} {{MHD}-limits to plasma confinement},\ }\href
  {https://doi.org/10.1088/0741-3335/26/1a/319} {\bibfield  {journal} {\bibinfo
   {journal} {Plasma Phys. Control. Fusion}\ }\textbf {\bibinfo {volume}
  {26}},\ \bibinfo {pages} {209} (\bibinfo {year} {1984})}\BibitemShut
  {NoStop}%
\bibitem [{\citenamefont {Connor}\ \emph {et~al.}(1998)\citenamefont {Connor},
  \citenamefont {Hastie}, \citenamefont {Wilson},\ and\ \citenamefont
  {Miller}}]{connor:1998}%
  \BibitemOpen
  \bibfield  {author} {\bibinfo {author} {\bibfnamefont {J.~W.}\ \bibnamefont
  {Connor}}, \bibinfo {author} {\bibfnamefont {R.~J.}\ \bibnamefont {Hastie}},
  \bibinfo {author} {\bibfnamefont {H.~R.}\ \bibnamefont {Wilson}},\ and\
  \bibinfo {author} {\bibfnamefont {R.~L.}\ \bibnamefont {Miller}},\ }\bibfield
   {title} {\bibinfo {title} {Magnetohydrodynamic stability of tokamak edge
  plasmas},\ }\href {https://doi.org/10.1063/1.872956} {\bibfield  {journal}
  {\bibinfo  {journal} {Phys. Plasmas}\ }\textbf {\bibinfo {volume} {5}},\
  \bibinfo {pages} {2687} (\bibinfo {year} {1998})}\BibitemShut {NoStop}%
\bibitem [{\citenamefont {Ryutov}\ \emph {et~al.}(2011)\citenamefont {Ryutov},
  \citenamefont {Berk}, \citenamefont {Cohen}, \citenamefont {Molvik},\ and\
  \citenamefont {Simonen}}]{ryutov:2011}%
  \BibitemOpen
  \bibfield  {author} {\bibinfo {author} {\bibfnamefont {D.~D.}\ \bibnamefont
  {Ryutov}}, \bibinfo {author} {\bibfnamefont {H.~L.}\ \bibnamefont {Berk}},
  \bibinfo {author} {\bibfnamefont {B.~I.}\ \bibnamefont {Cohen}}, \bibinfo
  {author} {\bibfnamefont {A.~W.}\ \bibnamefont {Molvik}},\ and\ \bibinfo
  {author} {\bibfnamefont {T.~C.}\ \bibnamefont {Simonen}},\ }\bibfield
  {title} {\bibinfo {title} {Magneto-hydrodynamically stable axisymmetric
  mirrors},\ }\href {https://doi.org/10.1063/1.3624763} {\bibfield  {journal}
  {\bibinfo  {journal} {Phys. Plasmas}\ }\textbf {\bibinfo {volume} {18}},\
  \bibinfo {pages} {092301} (\bibinfo {year} {2011})}\BibitemShut {NoStop}%
\bibitem [{\citenamefont {Bernstein}\ \emph {et~al.}(1958)\citenamefont
  {Bernstein}, \citenamefont {Frieman}, \citenamefont {Kruskal}, \citenamefont
  {Kulsrud},\ and\ \citenamefont {Chandrasekhar}}]{bernstein:1958}%
  \BibitemOpen
  \bibfield  {author} {\bibinfo {author} {\bibfnamefont {I.~B.}\ \bibnamefont
  {Bernstein}}, \bibinfo {author} {\bibfnamefont {E.~A.}\ \bibnamefont
  {Frieman}}, \bibinfo {author} {\bibfnamefont {M.~D.}\ \bibnamefont
  {Kruskal}}, \bibinfo {author} {\bibfnamefont {R.~M.}\ \bibnamefont
  {Kulsrud}},\ and\ \bibinfo {author} {\bibfnamefont {S.}~\bibnamefont
  {Chandrasekhar}},\ }\bibfield  {title} {\bibinfo {title} {An energy principle
  for hydromagnetic stability problems},\ }\href
  {https://doi.org/10.1098/rspa.1958.0023} {\bibfield  {journal} {\bibinfo
  {journal} {Proc. R. Soc. Lond. A}\ }\textbf {\bibinfo {volume} {244}},\
  \bibinfo {pages} {17} (\bibinfo {year} {1958})}\BibitemShut {NoStop}%
\bibitem [{\citenamefont {Cheng}\ and\ \citenamefont
  {Chance}(1987)}]{cheng:1987}%
  \BibitemOpen
  \bibfield  {author} {\bibinfo {author} {\bibfnamefont {C.}~\bibnamefont
  {Cheng}}\ and\ \bibinfo {author} {\bibfnamefont {M.}~\bibnamefont {Chance}},\
  }\bibfield  {title} {\bibinfo {title} {Nova: A nonvariational code for
  solving the mhd stability of axisymmetric toroidal plasmas},\ }\href
  {https://doi.org/https://doi.org/10.1016/0021-9991(87)90023-4} {\bibfield
  {journal} {\bibinfo  {journal} {J. Comput. Phys.}\ }\textbf {\bibinfo
  {volume} {71}},\ \bibinfo {pages} {124 } (\bibinfo {year}
  {1987})}\BibitemShut {NoStop}%
\end{thebibliography}


\end{document}